# High Rate Hybrid MnO$_2$@CNT Fabric Anode for Li-Ion Batteries: Properties and Lithium Storage Mechanism by In-Situ Synchrotron X-Ray Scattering

*Moumita Rana,[a] Venkata Sai Avvaru,[b] Nicola Boaretto,[a,c] Víctor A. de la Peña O'Shea,[d]*

*Rebeca Marcilla,[c] Vinodkumar Etacheri\*[a] and Juan J. Vilatela\*[a]*

[a]IMDEA Materials, Eric Kandel 2, 28906 Getafe, Madrid, Spain

[b]Faculty of Science, Autonoma University of Madrid, C/ Francisco Tomás y Valiente, 7, Madrid 28049, Spain

[c]IMDEA Energy, Avda. Ramón de la Sagra 3, 28935 Móstoles, Madrid, Spain

[d.]Photoactivated Processes Unit, IMDEA Energy Institute, Avda. Ramón de la Sagra 3, Parque Tecnológico de Móstoles, 28935 Móstoles, Madrid, Spain.

\* vinodkumar.etacheri@imdea.org, juanjose.vilatela@imdea.org

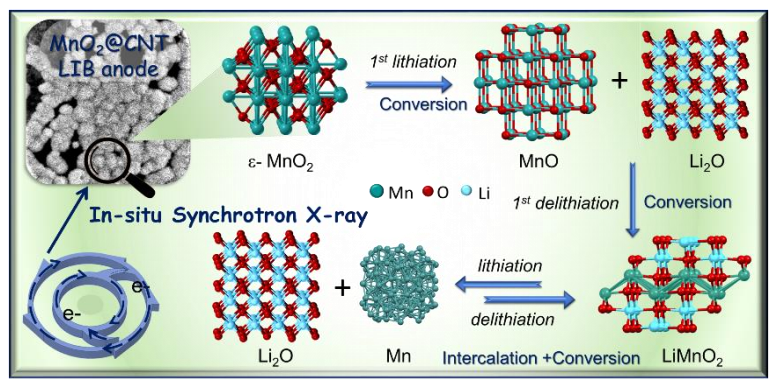

**Abstract:** High-performance anodes for rechargeable Li-ion battery are produced by nanostructuring of the transition metal oxides on a conductive support. Here, we demonstrate a hybrid material of MnO2 directly grown onto fabrics of carbon nanotube fibres, which exhibits notable specific capacity over 1100 and 500 mAh/g at a discharge current density of 25 mA/g and 5 A/g, respectively, with coulombic efficiency of 97.5 %. Combined with 97 % capacity retention after 1500 cycles at a current density of 5 A/g, both capacity and stability are significantly above literature data. Detailed investigations involving electrochemical and in situ synchrotron X-ray scattering study reveal that during galvanostatic cycling, MnO2 undergoes an irreversible phase transition to LiMnO2, which stores lithium through an intercalation process, followed by conversion mechanism and pseudocapacitive processes. This mechanism is further confirmed by Raman spectroscopy and X-ray photoelectron spectroscopy. The fraction of pseudocapacitive charge storage ranges from 27% to 83%, for current densities from 25 mA/g to 5 A/g. Firm attachment of the active material to the built-in current collector makes the electrodes flexible and mechanically robust, and ensures that the low charge transfer resistance and the high electrode surface area remain after irreversible phase transition of the active material and extensive cycling.

1. **Introduction:**

Lithium ion battery (LIB) is a crucial technology for most envisaged renewable energy schemes. Owing to their high energy density, wide voltage range they can be integrated in applications ranging from portable electronics to large-scale grids. [1,2] In the context of the electric mobility, LIBs have widespread use in automobiles and are the benchmark of current efforts for electric aircraft.[3,4] In electric transport, in addition to the ever-increasing need for higher specific energy density, there is particularly strong interest in performance at high rates in order to enable fast charge/discharge,[5–7] and in new battery concepts with augmented mechanical properties that provide: improved fatigue resistance and flexibility [8–10], enabled integration into existing structural components, use as load-bearing elements (i.e. structural batteries).[7,11,12]

Transition metal oxides (TMO) have shown promising activity as conversion materials for LIB.[13] Among them, manganese oxides stand out due to their high theoretical capacity (e.g. MnO, $Mn_3O_4$, $Mn_2O_3$, and $MnO_2$ have theoretical capacity of 756, 937, 1018, and 1230 mAh $g^{-1}$ respectively)[14] and a lower operating potential compared to Fe-, Co-, and Ni based transition metal oxides.[15,16] Nano-structuring of TMOs increases their high surface area, which in turn results in significantly high specific capacity from pseudocapacitive processes. This "hybrid" energy storage mode produces high capacity at high current densities.[17,18] Maier and coworkers, for example, have demonstrated a 60% enhancement in specific capacity in nanoporous anatase $TiO_2$ compared to the same bulk material, when used as LIB anode.[19]

However, there are several challenges associated with TMO electrodes for LIB that prevent us from exploiting their full potential. From the point of view of fabrication, it is straightforward to deposit nanostructured TMOs on a conventional metallic current collector (e.g. Al, Cu), but the resulting interface is very poor and cannot withstand the large structural (volume) change in TMO during charge/discharge cycles, leading to agglomeration and sintering of nanostructures.[20–23] This agglomeration can significantly reduce capacity due to a lower surface for pseudocapacitive processes and slower lithium diffusion kinetics. A more promising strategy is to stabilise the TMO onto a porous network of nanocarbons, either carbon nanotubes (CNT) or graphene.[1,24–29] The synthetic strategies for Lithium ion battery (LIB) is a crucial technology for most envisaged renewable energy schemes. Owing to their high energy density, wide voltage range they can be integrated in applications ranging from portable electronics to large-scale grids.[1,30,31] In the context of the electric mobility, LIBs have widespread use in automobiles and are the benchmark of current efforts for electric aircraft.[3,4] In electric transport, in addition to the ever-increasing need for higher specific energy density, there is particularly strong interest in performance at high rates in order to enable fast charge/discharge.[32–34] Moreover in new battery concepts with augmented mechanical properties are gathering huge attention since they provide: improved fatigue resistance and flexibility,[8–10,35] enabled integration into existing structural components, and use as load-bearing elements (i.e. structural batteries).[11,12]

Transition metal oxides (TMO) have shown promising activity as conversion materials for LIB.[13] Among them, manganese oxides stand out due to their high theoretical capacity (e.g. MnO, $Mn_3O_4$, $Mn_2O_3$, and $MnO_2$ have theoretical capacity of 756, 937, 1018, and 1230 mAh $g^{-1}$ respectively)[14] and a lower operating potential compared to Fe-, Co-, and Ni based transition metal oxides.[15,16] Nano-structuring of TMOs increases their specific surface area, which in turn results in significantly high specific capacity from pseudocapacitive processes.[36] This "hybrid" energy storage mode produces high capacity at high current densities.[18,37] Such strategy also leads to substantial improvement in power performance and cycling behaviour due to shorter ion diffusion paths and facilitating Li insertion with minimized internal strain.[14,38] Maier and coworkers, for example, have demonstrated a 60% enhancement in specific capacity in nanoporous anatase $TiO_2$ compared to the same bulk material, when used as LIB anode.[19] For manganese dioxide, α-$MnO_2$ nanorods demonstrated significantly enhanced electrochemical performance with initial reversible capacity of 1075 mAh $g^{-1}$ at current density of 0.1 A $g^{-1}$,[39] and a 3D δ-$MnO_2$ mesopore nanostructure delivered a high lithium storage capacity of 905 mAh $g^{-1}$ at 0.1 A $g^{-1}$ with a capacity retention of 94% after 200 cycles at 1 A $g^{-1}$.[18]

However, there are several challenges associated with TMO electrodes for LIB that prevent us from exploiting their full potential. From the point of view of fabrication, it is straightforward to deposit nanostructured TMOs on a conventional metallic current collector (e.g. Al, Cu), but the resulting interface is very poor and cannot withstand the large structural (volume) change in TMO during charge/discharge cycles, leading to agglomeration and sintering of nanostructures.[20–23] This agglomeration can significantly reduce capacity due to a lower specific surface area for pseudocapacitive processes and slower lithium diffusion kinetics. A more promising strategy is to stabilise the TMO onto a porous network of nanocarbons, either carbon nanotubes (CNT) or graphene.[1,24–29] The synthetic strategies for the preparation of CNT fiber based hybrids mostly rely on the in situ growth of nanocrystals on the CNTF by electrodeposition, sol-gel, electroless deposition, solvothermal and high temperature vapour deposition methods.[26,40] Such strategies not only result in an electrochemically stable interface between metal oxide and carbon support, but also helps to avoid the use of polymeric binders during electrode preparation. As an example, nanostructured $MnO_2$ directly grown onto unidirectional fabrics of CNT fibres resulted in pseudocapacitive electrodes with high capacitance, stability above 6000 cycles under high potential, and large mechanical robustness, stemming from the strong adhesion of the TMO to the built-in CNT fabric current collector.[41]

More importantly, stabilising nanostructured TMOs onto porous CNT current collectors can enable a detailed study on their structural changes when used as LIB electrodes, a necessary step towards further improvements in performance and cyclability. The vast number of works on performance of $MnO_2$ LIB anodes contrasts with very few studies on the mechanism of lithium storage in $MnO_2$ anodes. Specifically, when TMOs are employed as anodes, the conversion process during charge-discharge is often assumed to be a reversible transformation of the parent TMO to $Li_2O$ and metal.[42,43] On the other hand, experimental studies reveal a far more complex process because of the various possible phase transformations and microstructural changes of the TMO. Using electron energy loss spectroscopy and selected area electron diffraction analysis, Chen *et al*. proposed that $\beta$-$MnO_2$ is only partially reduced to $LiMn_3O_4$, instead of being reduced to metallic Mn.[44] From ex-situ X-ray diffraction study, Fang *et al*. found that lithiation of $MnO_2$ results in $Li_xMnO_2$ (x=0.96) and $Li_2MnO_2$, which is further reduced to metallic Mn and $Li_2O$ at 0 V,[45] consistent with observations by other groups.[46] However, after complete delithiation at 3.0 V, MnO was found to be the end product.[45] Such permeant reduction of $Mn^{4+}$ to lower oxidation state (+2) during the charge-discharge cycle reduces specific capacity after cycling. Contrary to this experimental evidence of permanent phase transformation during charge/discharge, it is common to find the widespread assumption of a reversible conversion of $MnO_2$ based solely on voltammetric profiles.[39,47]

Herein, we perform in situ synchrotron X-ray scattering measurements during lithium storage in the hexagonal phase (Akhtenskite) of $MnO_2$. We synthesized free-standing $\varepsilon$-$MnO_2$-CNT fibre hybrids ($MnO_2$@CNTF) by producing a uniform coating of porous $MnO_2$ flower-like nanostructures around bundles of CNT. When used as an anode in half-cell configuration against Li-metal, this material shows specific capacity of 1153, 519 and 344 mAh/g at scan rates of 25 mA/g, 5 and 10 A/g respectively, a rate capability above that of most related materials. At low scan rate, the lithium storage process occurs through a conversion mechanism, whereas at high scan rate the mechanism mostly relies on pseudocapacitive Li storage. In situ synchrotron X-ray scattering study reveals that during the first charge-discharge cycle, the hexagonal structure of parent $MnO_2$ slowly converts to lithiated manganese oxide. In the consecutive galvanostatic cycles, this structure undergoes a reversible phase transformation with $Li_2O$ and Mn. During this process the porous $MnO_2$ structures experience 'electrochemical milling', forming a nanoscopic thin oxide coating around the CNT bundles that leads up to 200% increase in the specific capacity of the material.

**Experimental Details**

**Chemicals**

Butanol, Ferrocene. Sodium nitrate (99.99%, Sigma Aldrich), Manganese nitrate (99.9%, Merck), Li metal sheet, 1 M $LiPF_6$ in ethylene carbonate and ethyl methyl carbonate (1:1-v/v, Sigma Aldrich), Manganese oxide ($\geq$99 %, Sigma Aldrich). All chemicals were used as received, without further purification.

**Synthesis of free standing CNT fiber veils**

The CNT fiber (CNTF) veils were prepared by chemical vapour deposition method at 1250 °C under hydrogen atmosphere in a vertical tube furnace using butanol, ferrocene and thiophene as carbon source, catalyst and promoter respectively.[48] The CNTF veil electrodes were prepared by winding the CNT fibers around a paper spool for 30 minutes, which was later taken out from the spool as free standing electrode. The aerial density of the samples was estimated to be ~ 0.7 mg/cm$^2$. The CNT fibres used in this work have a specific surface area ~250 m$^2$/g and longitudinal conductivity around 3.5×105 S/m.[49]

**Electrodeposition of MnO$_2$ on CNTF**

The free-standing CNT fiber veils were electrochemically oxidized by using chronoamperommetry at a constant voltage of 2.5 V for 1 minute in 0.1 M Na$_2$SO$_4$.[50] The functionalized CNTF veils were rinsed with water and subjected to electrodeposition using a solution of NaNO$_3$ (255mg) and Mn(NO$_3$)$_2$ (600 mg) in water-ethanol (30 mL, 1:1) mixture as electrolyte. The electrodes were produced under chronopotentiometric conditions with a constant current density of 300 µA/cm$^2$ for 1 hour, followed by thorough washing with water and ethanol. The electrodeposited samples were annealed in air at 120 °C for over-night. The mass fraction of MnO$_2$ can be adjusted by varying electrodeposition parameters. The conditions used here were selected to yield a mass fraction in the range 40-65%. To estimate the weight saving factor, we prepared a reference material by coating a dispersion of commercial MnO$_2$, acetylene black, PVDF (7:2:1) on a Cu foil. The electrode was dried in oven at 80 °C under vacuum for overnight and used for electrochemical measurements.

**Characterizations**

The samples were characterized using field effect scanning electron microscopy (FESEM, FEI Helios NanoLab 600i), transmission electron microscopy (TEM, Talos F200X FEG, 200 kV), selected area electron diffraction (SAED), high-angle annular dark-field imaging- energy-dispersive X-ray spectroscopy (HAADF-EDS), Raman spectroscopy (Ranishaw, fitted with a 532 nm laser source), thermogravimetric analysis (TGA, Q50, TA Instruments), Brunauer-Emmett-Teller (BET) adsorption (Micromeritics, TriStar II Plus Version 3.01), wide angle X-ray scattering (WAXS, CDD detector LX255-HS, Rayonix) and Near Ambient Pressure X-ray photoelectron spectroscopy (NAP-XPS SPECS System, Al Kα monochromated source and 1-DLD detector).

**Electrochemical Measurements**

The samples were tested as working electrode in coin cell configuration (type 2032). Li metal plates were used as reference and counter electrode, glass fiber (Whatmann) as separator, and 1 M LiPF$_6$ solution in a mixture of EC-EMC (1: 1 v/v) was used as electrolyte. The coin cells were assembled in glove box, filled with argon (concentration of $O_2$ and $H_2O$ < 1.0 ppm). The charge storage behaviour of the sample was characterized by charge-discharge (CD) using Neware battery testing unit, cyclic voltammetry (CV) and electrochemical impedance spectroscopy (EIS) using Biologic SP-200 electrochemical workstation. The voltages mentioned throughout the manuscript is with respect to Li/Li$^+$ redox potential.

**In situ measurements**

In situ synchrotron WAXS measurements were performed using a home-made gas-tight teflon cell, fitted with two kapton windows. The sample was placed at the centre of the windows and connected to the external cables through metallic tape. The electrochemical measurements were performed in a two electrode configuration. The experiments were performed at the Non Crystalline Diffraction (NCD-SWEET) beamline 11 of ALBA Synchrotron Light Facilities. The WAXS patterns were calibrated using chromium (III) oxide ($Cr_2O_3$) EAXS pattern as reference. Radial profiles were obtained after azimuthal integration of scattering intensity.

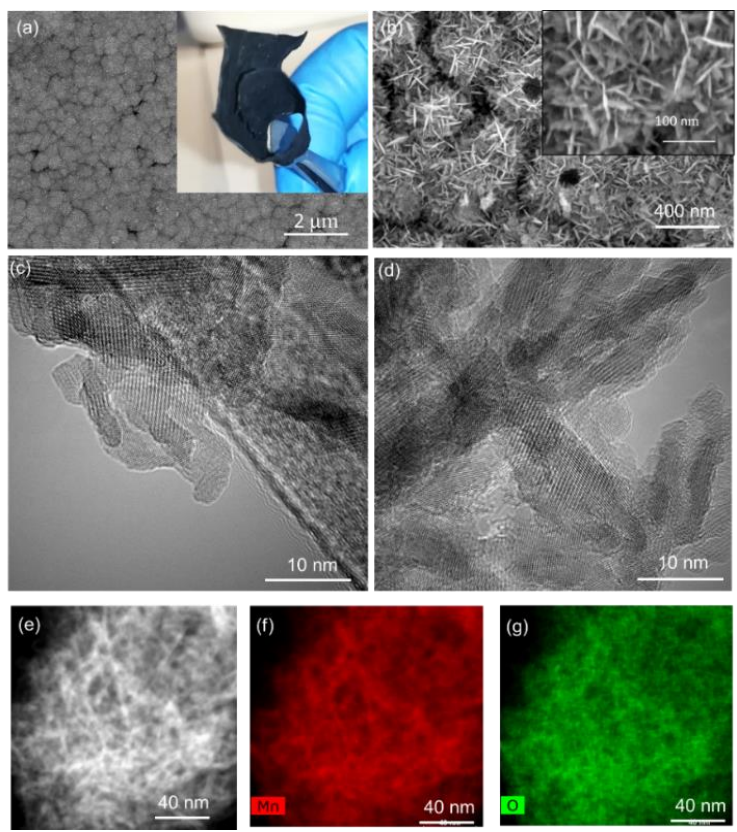

*Figure 1.* *(a, b) FESEM images of MnO2@CNT showing uniform growth of nanostructured hollow MnO$_2$ flowers on CNT fiber veils. Inset in (a) shows the flexible nature of macroscopic MnO$_2$@CNT hybrid. Inset in (b) shows high-resolution image of interconnected nanopetals. (c, d) TEM images of the hybrid showing the growth of polycrystalline nanopetals of MnO$_2$ on functionalized CNT. (e) HAADF and (f, g) EDS elemental mapping images corresponding to elements Mn and O respectively of a MnO$_2$ flower.*

**Results and Discussion:**

Free-standing hybrid MnO$_2$@CNT fabrics were prepared by electrodepositing manganese oxide on electrochemically pre-functionalized CNT fiber fabrics from a solution of manganese nitrate in water and ethanol mixture (1:1).[41,50] The FESEM images of the sample at different magnifications are shown in **Figure 1a-d**. The inset in **Figure 1a** shows the flexible nature of the macroscopic electrode sample, which can withstand rolling up to a radius < 8 mm. The CNTF is uniformly covered by the porous flower-like nanostructures with average diameter of 550 nm (±100), comprised of interconnected thin sheets with thickness of ~13 nm (± 6) and length of 50-100 nm (**Figure 1a, b** and **Figure S1**), resembling petals.

These nanocrystals are predominantly aligned with their main axis perpendicular to the CNT bundles and aggregated into semi-spherical structures, as shown in **Figure 1b and S1e**. The uniform growth of the manganese oxide nanostructures on CNTF can be related to the mild pre-oxidation of the CNT fibers by electrochemical functionalization, as the same electrodeposition treatment on the pristine CNT samples leads to only partial coverage of the fibers by larger manganese oxide porous structures, as shown in **Figure S2**. The oxygen functional groups at the CNT surface apparently act as nucleation sites for the deposition of metal oxide during chronopotentiometric process, thereby facilitating uniform growth of the smaller size flowers. As key features of this type of electrodes we highlight the large size, strength and low electrical resistance of the interface between MnO$_2$ and the CNT bundles. For the samples in this study, no evidence of detachment of MnO$_2$ has been observed after extensive manipulation and repeated bending, in agreement with recent reports on related hybrids used for LIB cathodes and which combine exceptional mechanical toughness and a 29 % increase in specific capacity compared to commercial LIB cathodes, preserved even after tensile fracture.[51]

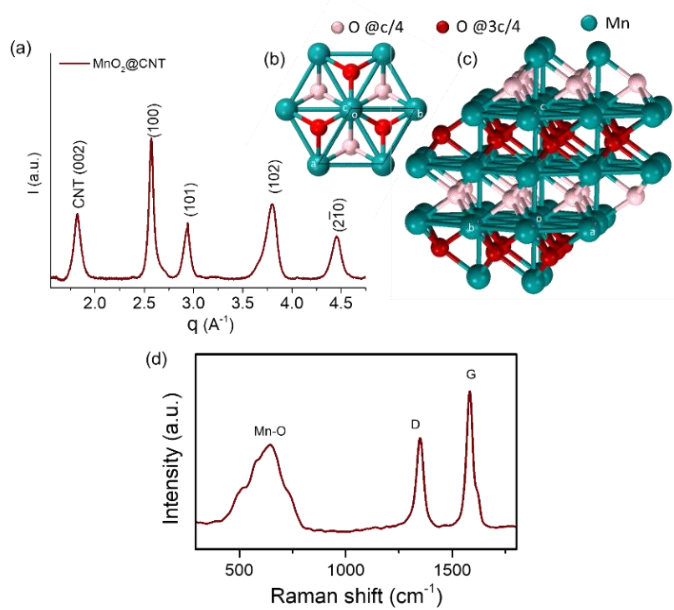

*Figure 2. (a) Synchrotron X-ray diffraction pattern of the MnO$_2$@CNTF hybrid. (b, c) Crystal structure of the hexagonal Akhtenskite phase of MnO$_2$ along (b) and perpendicular (c) to the c-axis. (d) Raman spectra of MnO$_2$@CNTF hybrid.*

**Figure 1c and d** show the high resolution TEM images of the $MnO_2$@CNTF hybrid. Under TEM, the "petals" of the manganese oxide appear as polycrystalline layer of manganese oxide. In **Figure 1c,** the high-resolution image of the CNT- manganese oxide interface reveals that, these nanocrystals are directly attached to the CNT surface. To investigate the elemental composition of this hybrid, we performed HAADF-EDS mapping. **Figure 1e-g** shows the HAADF image and elemental mapping of the same particle corresponding to manganese and oxygen. BET isotherm analysis was performed to estimate surface area of the $MnO_2$@CNTF hybrid. **Figure S3b** shows isothermal adsorption and desorption profiles of $N_2$ at 77K, which can be assigned as type II. The surface area of the $MnO_2$@CNTF was found to be 130 $m^2/g$.

In order to understand the crystal structure, we have acquired the synchrotron XRD pattern of the $MnO_2$@CNTF hybrid, as shown in **Figure 2a**, which can be indexed on the basis of the hexagonal Akhtenskite phase of $MnO_2$ with lattice parameters a = b = 2.829 Å, c= 4.410 Å (c/a= 1.559) and α=β= 90° and γ= 120° with a space group of $P6_3/mmc$.[52,53] In **Figure 2b and c** the crystal structure of this phase is shown along and perpendicular to c-axis respectively. In this structure, the $Mn^{4+}$ ions occupy the corners as well as centre of the edges at ½ c distance, and the hexa-coordinated $O^{2-}$ ions occupy alternate octahedral sites.

**Figure 2d** shows the Raman spectra of $MnO_2$@CNTF hybrid. The two sharp peaks at 1352.8, 1582.3 $cm^{-1}$ can be assigned to D and G bands of oxygen functionalized CNT fibers respectively, which are originated from disordered features of CNT and tangential vibrational mode displacement (E2g mode) respectively. Here the $I_D/I_G$ ratio of the functionalized CNT was found to be 0.75. The peaks at 645, 583 and 530 $cm^{-1}$ can be attributed to the symmetric stretching and bending modes of Mn-O in [$MnO_6$] octahedra.[54] The high intensity of such $MnO_2$ peaks confirms the extensive coverage of the CNTF by crystalline $MnO_2$.[55,56] To estimate the loading of $MnO_2$ in the hybrid, we performed thermogravimetric analysis in aerial atmosphere and calculated mass fraction taking into account the catalytic combustion of CNT in presence of chemically attached Mn-O bonds.[57] (**Figure S4).** The sample analysed in this work has a mass fraction of $MnO_2$ of 55 wt.%, which we found to provide a good balance between electrochemical properties and mechanical robustness.

**Electrochemical performance**

The lithium storage capacity of $MnO_2$@CNTF hybrid was estimated using a coin cell configuration with Li foil as counter electrode. The open circuit voltage (OCV) of the pristine cell was measured to be 3 V (*vs*. $Li/Li^+$). **Figure 3a** shows the first two cyclic voltammograms at a scan rate of 0.1 mV/s. The initial cathodic profile (lithiation) contains a strong peak at 0.12 V corresponding to conversion of $Mn^{4+}$ to metallic Mn.[58] An additional cathodic peak at 0.65 V with a small hump at 0.78 V can be attributed to the formation of solid electrolyte interphase (SEI).[59] In the following anodic cycle (delithiation), two peaks were observed at 1.27 and 2.19 V, which can be attributed to the oxidation of $Mn^0$ to $Mn^{2+}$ and $Mn^{2+}$ to higher oxidation states of Mn ($Mn^{3+}/Mn^{4+}$) repectively.[58] Interestingly, in the 2$^{nd}$ cathodic profile the cathodic peaks appeared at 0.31 and 0.20 V, which are distinctly different from the first cycle. Such higher lithiation voltage of $MnO_2$ after the first cycle is indicative of an irreversible structural change of the oxide layer, taking place during the first redox cycle.[45,60] In the 2$^{nd}$ delithiation process the anodic peaks were observed almost at the same position as that of the first one (1.31 and 2.21 V). For reference, we performed cyclic voltammetry using pristine CNT as anode under identical conditions, which was found to be distinctively different from that of $MnO_2$@CNTF hybrid (**Figure S5a**).

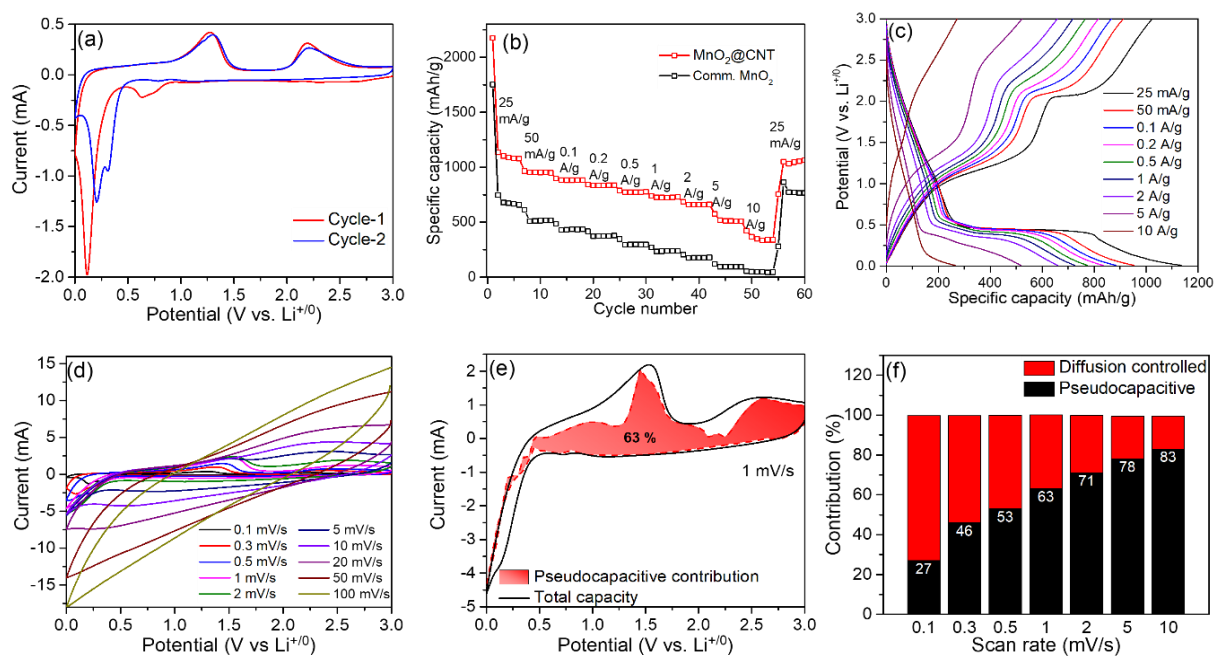

*Figure 3*. *(a) First two cyclic voltammograms of $MnO_2$@CNTF hybrid at a scan rate of 0.1 mV/s. (b) Rate profiles of $MnO_2$@CNTF hybrid and commercial $MnO_2$ at current densities of 25 mA/g to 10 A/g. (c) Voltage profiles of $MnO_2$@CNTF hybrid corresponding to its rate profile in (b). (d) cyclic voltammograms of $MnO_2$@CNTF hybrid at different scan rates. (e) Deconvoluted contribution of pseudocapacitive processes in the cyclic voltammograms at a scan rate of 1 mV/s. (e) Estimated contribution of diffusion controlled and pseudocapacitive reactions with different scan rates.*

The redox features of $MnO_2$@CNTF hybrid from cyclic voltammograms are in complete accordance with the galvanostatic charge-discharge (CD) profiles. The discharge capacity values of MnO$_2$@CNTF at current densities 25 mA/g to 10 A/g are shown in **Figure 3b. Figure S5b** shows the first three CD profiles of the MnO$_2$@CNTF at current density of 25 mA/g. In the first discharge process (lithiation), the specific capacity of MnO$_2$@CNTF hybrid was found to be 2175 mAh/g, dropping to 1153 mAh/g after the second cycle and remain fairly stable afterwards. Such decrease in the specific capacity in the first cycle can be attributed to the irreversible consumption of lithium in the formation of SEI layer as well as partial lithiation of manganese oxide (*vide infra*). The specific capacities of MnO$_2$@CNTF at current densities of 0.1, 1,5 and 10 A/g were found to be 882, 722, 519 and 344 mAh/g respectively. Even though the transition metal oxides are well known for their high Li-ion storage capacity at low current density, such a high capacity at high current density is quite impressive for these materials. For reference, we compared the rate profile of commercial MnO$_2$ in Figure 3b (and Figure S6). The specific capacity values were estimated to be 680, 434, 236 and 93 mAh/g at current densities of 0.025, 0.1, 1 and 5A/g respectively, which are much lower with respect to the performance of our MnO$_2$@CNTF hybrid. When compared with the literature reports on high performance Li-ion storage material, this MnO$_2$@CNTF hybrid exhibits performance well above average for all current densities, and to our knowledge, the highest value reported at high current density for MnO$_2$ as starting anode material. (see the comparison **Table ST1** and **Figure S5d** in ESI).

The voltage profiles of the material at various current densities are shown in **Figure 3c**. For each particular current density, the voltage profiles were found to be near identical, with coulombic efficiency > 97.5%. This confirms that after the irreversible transformation in the first cycle, subsequent charge-discharge cycles are fairly reversible. We also note that the redox features of these CD profiles are significantly different from the same of pristine CNT fiber electrode (**Figure S5a**), confirming the predominant contribution of MnO$_2$ in the charge storage mechanism.

The voltage profiles of the lithiation process at lower current densities have three distinct regions: (i) an almost linear voltage decrease from 3V to 0.45 V, (ii) a flat plateau around 0.45 V, followed by (iii) a sloping profile up to complete lithiation. The plateau in region (ii) is representative of the conversion reaction of manganese oxide, whereas the presence of the sloping profiles in regions (i) and (iii) are a signature of pseudocapacitive Li storage at the inorganic material surface and at newly formed Li$_2$O –metal interface, respectively.[61] With increasing current density, the plateau-like feature (region ii) decreases and finally merges with the region (iii). In order to estimate the contribution of conversion reaction and pseudocapacitive processes towards total Li-ion storage capacity at different current densities, we performed CVs at different scan rates (**Figure 3d**). When the scan rate is higher than 5 mV/s, the redox peaks are hardly distinguishable, thus indicating negligible contribution from diffusion controlled conversion reactions. We calculated the contribution of capacity from diffusion controlled conversion reactions and pseudocapacitive surface reactions using the following equation.[36,62,63]

$$I(V) = K_1\upsilon + K_2\upsilon^{1/2} \quad \text{(eq. 1)}$$

Where $\upsilon$ is the scan rate, and $K_1$ and $K_2$ are scan rate independent constants. $K_1\upsilon$ stands for capacitive contribution and $K_2\upsilon^{1/2}$ denotes diffusive contribution. As shown in **Figure 3e**, at a scan rate of 0.1 mV/s (nearly equivalent to a current density of 100 mA/g), the contribution of the diffusion-controlled reaction (i.e. conversion reaction) is as high as 73%, which decreases to 17% at a high scan rate of 10 mV/s (comparable to lithiation in galvanostatic control at current density of 5 A g$^{-1}$, **Figure 3f**). This shows that at low current density, the charge storage mechanism mostly occurs through conversion process, whereas at high current densities, it is based on pseudocapacitive surface reactions only.

The high rate performance of MnO$_2$@CNTF can be mostly ascribed to high surface area of the oxide layer around the CNTF, which itself has a high specific surface area (SSA) of 250 m$^2$/g.[49] In addition to templating the growth of the metal oxide, the highly conducting porous CNT fibre support can buffer the structural change of MnO$_2$ active material during the conversion reactions, which would otherwise undergo irreversible agglomeration, thereby significantly compromising on the charge storage capacity in the following cycles and producing unstable rate profiles.[64]

**Lithium storage mechanism in ε-MnO$_2$:**

To understand the structural changes of manganese dioxide layer during the lithiation/delithiation process, we performed in situ synchrotron WAXS measurements using a gas-tight teflon cell, fitted with two kapton windows, built in-house (**Figure S7**). The OCV of the pristine device was recorded to be 3V. The charge-discharge profiles during the in situ measurements are in accordance with the redox signature from the cyclic voltammogramms (CV) at scan rate of 0.1 mV/s. **Figure 4a, b** show the voltage profiles of 1$^{st}$ and 2$^{nd}$ charge-discharge cycles respectively, and **Figure 4c, d** show the corresponding in situ WAXS patterns at different voltages during the discharge and charge process. The WAXS pattern of the sample at OCV matches exactly the pattern of hexagonal MnO$_2$. During the lithiation and delithiation process, clear phase change of the oxide is observed, which is not reversible during the first cycle (**Figure 4c**). The phase obtained at the end of the first CD cycle is distinctively different from the original pattern of the pristine MnO$_2$. This newly formed phase undergoes a reversible transformation in the 2$^{nd}$ galvanostatic cycle (**Figure 4d**). The simulated position of the peaks of various materials along with appropriate references are summarized in **Table ST2** in the supporting information.

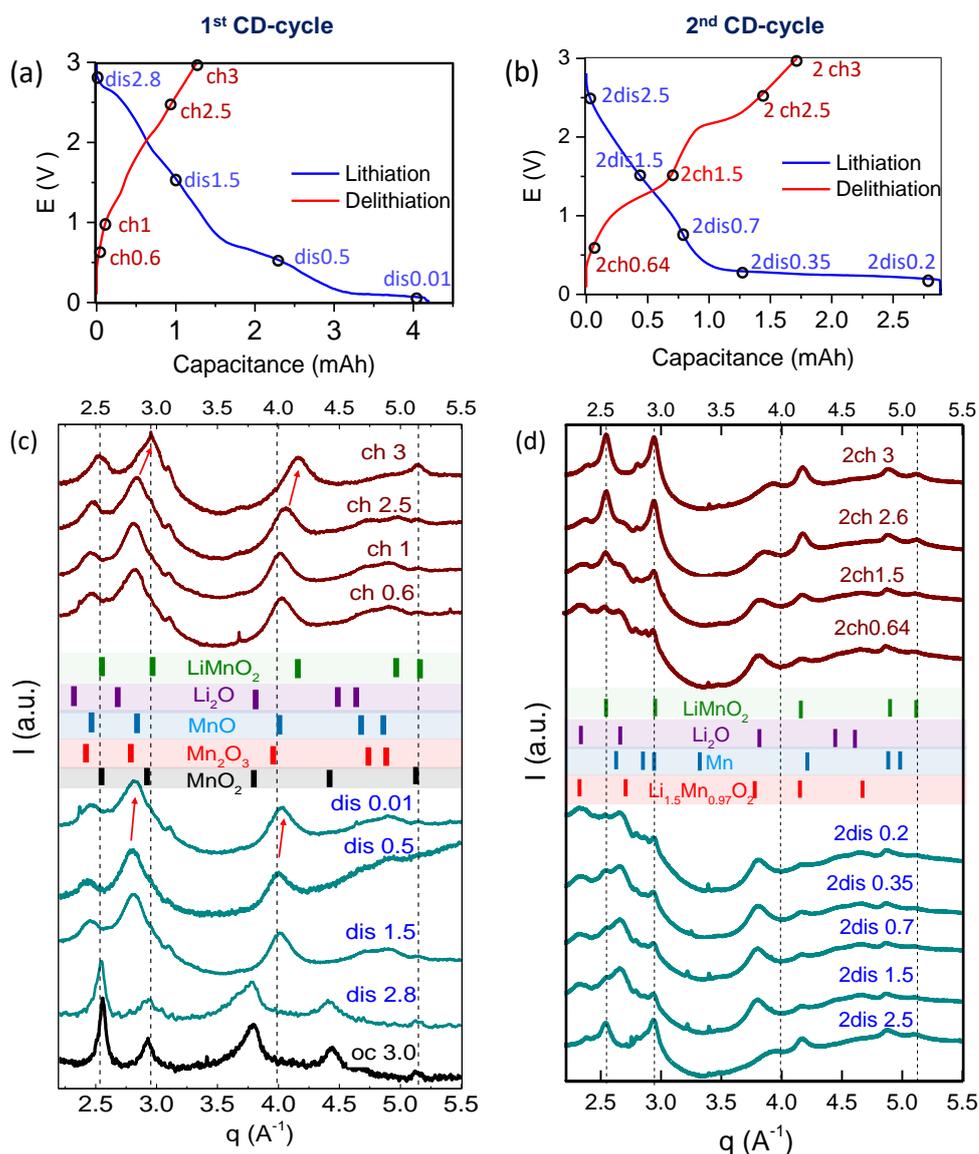

*Figure 4. The voltage profiles (a, b) and corresponding in situ synchrotron X-ray scattering patterns (WAXS) (c, d) for the lithiation and delithiation process of 1st (a, c) and 2nd cycle (b, d) at a current density of 100 mA/g. In Figure c and d, the lithiation and delithiation process of 1st (a, c) and 2nd cycle (b, d) at a current density of 100 mA/g. In Figure c and d, the vertical lines correspond to the simulated patterns of different materials from the crystal structure database (ICSD, ICDD-PDF2).*

The voltage profile of the first galvanostatic lithiation presents features of a pseudocapacitive process rather than an exclusive conversion process (**Figure 4a**). Part of the large irreversible capacity loss in the first cycle can be related to the formation of amorphous SEI layer on the high-surface area electrode. In addition, during the 1st discharge, as the voltage reaches 1.5 V, the peaks of hexagonal $MnO_2$ disappear and evolve to the orthorhombic phase of $Mn_2O_3$. When the voltage drops further to 0.01V (complete lithiation) with a plateau, small shifts in the WAXS peak positions towards higher '$q$' were observed (shown with arrow in **Figure 4c**), which can be assigned to the diffraction pattern of the cubic phase of MnO. Further broadening of the peak at 2.94 Å$^{-1}$ can be related to evolution of Li$_2$O Due to broad nature of the WAXS peaks of the manganese oxides, the peaks corresponding to Li$_2$O could be merged with the ones of manganese oxide peaks, thus increasing their broadening at the end of the lithiation process (2.44 and 2.94 Å$^{-1}$).

The process occurring during the first lithiation would thus consist of:

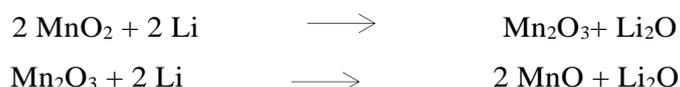

$$2\ MnO_2 + 2\ Li \longrightarrow Mn_2O_3 + Li_2O$$
$$Mn_2O_3 + 2\ Li \longrightarrow 2\ MnO + Li_2O$$

As shown in **Figure 4c**, with increasing voltage, the peaks corresponding to MnO slowly move toward higher q values. Considering the oxidation of the manganese center with gradual lithium extraction, the final WAXS pattern obtained at 3 V can be assigned to the orthorhombic LiMnO$_2$. Therefore, the redox process during the 1$^{st}$ charging can be expressed as:

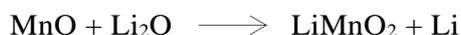

$$MnO + Li_2O \longrightarrow LiMnO_2 + Li$$

Unlike the charging profiles during the rate capability tests, in the first charging profile in the in situ WAXS measurement (delithiation process), there are hints of plateaus from the oxidation of metallic Mn to higher oxidation states. This implies that during the previous discharge process a small conversion of MnO$_2$ to metallic Mn and Li$_2$O cannot be neglected.

Interestingly, the voltage profile and the change in the in situ WAXS patterns during the 2$^{nd}$ galvanostatic cycle is distinctively different from the first one (**Figure 4b, d**). Here the discharge profile reaches a cathodic plateau at 0.4 V following a pseudocapacitive feature, whereas the charging profile has two distinct anodic signatures for a two-step oxidation process. During discharge, as the voltage reaches 1.5 V, the intensity of the peaks corresponding to LiMnO$_2$ decreases with the evolution of new sets of peaks that can be assigned to a more lithium rich manganese oxide with enhanced lattice spacing (Li$_{1.5}$Mn$_{0.97}$O$_2$), possibly through an intercalation process,[61] which is consistent with the electron uptake by the active material. Upon further lithiation, this intermediate reduces to metallic Mn and Li$_2$O. During consecutive charging, slow evolution to LiMnO$_2$ was observed, which might involve the intermediate lithiated form of manganese oxide. A schematic diagram of the observed structural transformations of MnO$_2$ from in situ WAXS measurement is shown in **Figure 5.** The associated reactions are also included for reference below.

Second discharge

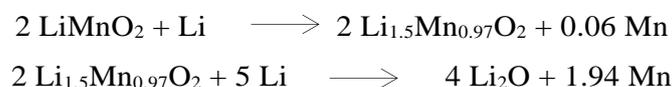

$$2\ LiMnO_2 + Li \longrightarrow 2\ Li_{1.5}Mn_{0.97}O_2 + 0.06\ Mn$$
$$2\ Li_{1.5}Mn_{0.97}O_2 + 5\ Li \longrightarrow 4\ Li_2O + 1.94\ Mn$$

Second charge

$$2\,Mn + 4Li_2O \longrightarrow 2\,LiMnO_2 + 6\,Li$$

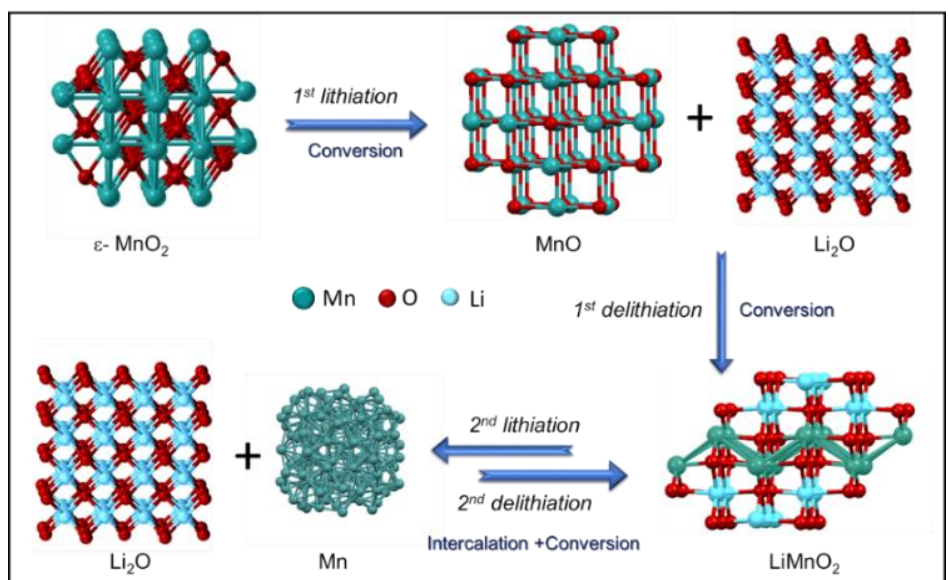

***Figure 5.*** *Schematic diagram for structural transformation of ε-MnO$_2$ observed by in situ WAXS measurements.*

These observations clarify the relation between Li storage mechanisms in MnO$_2$ anodes and the corresponding voltage profiles. Firstly, it can be inferred that even though MnO$_2$ was used as starting electroactive material, upon cycling in voltage window of 0-3V, the first conversion reaction is not reversible. Part of the irreversible capacity loss in the 1$^{st}$ cycle can be attributed to the irreversible lithium consumption by MnO$_2$. Indeed, at the end of several galvanostatic cycles, the sample was found to contain significant amount of LiMnO$_2$, rather than pristine MnO$_2$ structure (**Figure S8b**), which confirms that the newly formed LiMnO$_2$ has chemical reversibility upon electrochemical (EC) cycling. The same also can be inferred from the Raman spectrum of the EC cycled sample (**Figure S8c**). Apart from the signature D and G peaks from CNT, the strong peak at 600 cm$^{-1}$ can be attributed to the stretching mode of Mn-O bonds, which has experienced significant red shift compared to that of the pristine material (**Figure 4d**). This can be delineated to decreased oxidation state of Mn and thereby decrease in the Mn-O bond energy. The smaller and broader peaks at 418, 467 cm$^{-1}$ and 245, 291 cm$^{-1}$ can be related to the deformation vibrations of the Mn–O–Mn bonds and vibrational modes of LiO$_6$ octahedra respectively.[65–67] Furthermore, we confirmed the oxidation states of Mn in the samples after 1$^{st}$ complete lithiation and the following delithiation using X-ray photoelectron spectroscopy. **Figure S9** shows the high resolution XP spectra of Mn (3s) in the (a) pristine sample, (b) at 0 V (1$^{st}$ cycle) and (c) at 3V. Valence state of manganese ions was determined using the Galakhov *et al.* approach[68], based on the splitting of the Mn 3s core-level spectrum from the exchange coupling between the Mn 3s hole and Mn 3d electrons. The magnitude of this splitting is proportional to (2S + 1), where S is the local spin of the 3d electrons in the ground state. The differences in binding energy between these two peaks (□BE) were used to determine the average oxidation state of Mn at different states of lithiation, using values provided by Alvarez Galvan *et al* for reference.[69] Binding energy differences come out as 4.9, 5.8 and 5.3 eV, which correspond to (a) $Mn^{4+}$, (b) $Mn^{2.1+}$ and (c) $Mn^{3.1+}$ respectively. These results are in complete accordance with the findings from our in situ WAXS experiment. In this regard, it is noteworthy that when $Li_2MnO_3$ was used as anode material by Muhammad *et al.*, after 1$^{st}$ galvanostatic cyclic, it converted to cubic rock salt type $Li_{0.5}Mn_{0.5}O$ through the formation of $Mn_3O_4$ as intermediate.[70] In both the cases, +4 oxidation state of Mn reduces to +3 in the 1$^{st}$ CD cycle, which undergoes reversible conversion in the following cycles. Thus, the electrochemical reversibility of $Mn^{4+}$ in such oxide rich environment is questionable.

These results also indicate that the nanostructured surface of this hybrid is beneficial to store significant amount of lithium through an intercalation process by increasing lithium concentration in the oxide. A similar increase in lithium concentration has been observed in $MoO_3$ anode, [61] for example. But, to the best of our knowledge, the present work provides the first evidence of Li-ion storage by intercalation for manganese oxide-based Li-ion anodes. This also confirms that the charge storage mechanism for the region-(i) in the discharge profiles (*vide supra*) can be assigned to Li-ion storage by intercalation mechanism.

Finally, the in situ WAXS data during charge-discharge processes show a gradual reduction in relative intensity and broadening of the peaks, ascribed to progressive loss of crystallinity during the repeated conversion of the nanostructured inorganic phase.

**Electrochemical stability**

Equipped with the insights into the storage mechanism process, we evaluated the electrochemical stability of the hybrid material by galvanostatic cycling at different current densities. **Figure 6a, b and c** show the electrochemical stability of the $MnO_2$@CNTF/Li upon galvanostatic cycling at current densities of 0.5, 1 and 5 A/g for 150, 300 and 1500 CD cycles respectively. When cycled at lower current densities (0.5 and 1 A/g), the $MnO_2$/CNTF hybrid shows a slow increase in the capacity up to 200% Similar phenomena also have been observed when $Mn_3O_4$, [29,] $MnO_2$ [32] and $Cu_2O–Li_2O$ [55] were used as electroactive materials. During galvanostatic cycling at 0.5 A/g, the specific capacity gradually increases to 152% of its initial value. Similar enhancement in the capacitance is observed for galvanostatic cycling at 1 A/g current density; after 220 cycles an enhancement of 187% in specific capacitance is observed. After reaching the highest value, the capacity remains constant during rest of the cycling test. Throughout the CD cycling the coulombic efficiency remain around 100% (average 99.3%).

Very importantly, the specific capacity value at 5 A/g current density remains at 97 % of its original value, with near unity coulombic efficiency even after 1500 galvanostatic cycles. The electrochemical stability of this MnO$_2$@CNTF hybrid in half-cell configuration is comparable and even superior to some of the literature reports.[18,42,43,71,72] This can be attributed to the presence of a stable interface between manganese oxide nanocrystals and carbon nanotubes, which not only buffers the structural change in the oxide layer, but also prevent agglomeration of the nanostructures during the charge–discharge process.

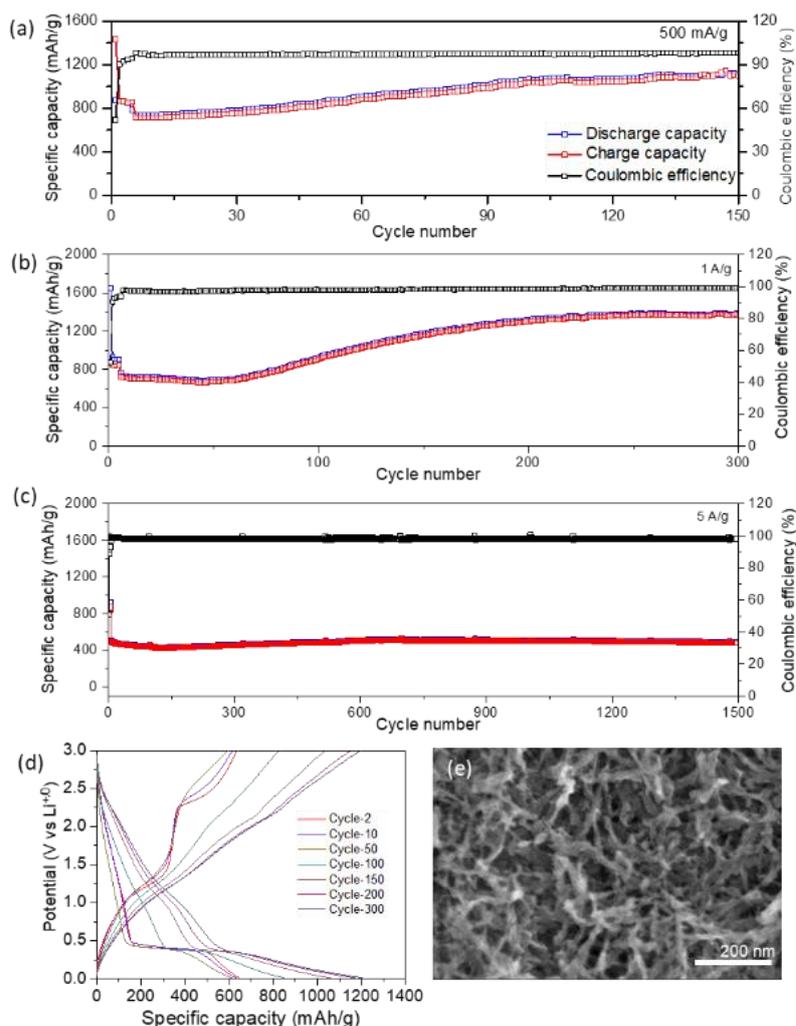

*Figure 6. Stability of MnO2@CNTF hybrid electrode over charge-discharge process of (a) 150 cycles at a current density of 0.5 A/g, (b) 300 cycles at a current density of 1A/g and (c) 1500 cycles at a current density of 5A/g. (d) The voltage profiles corresponding to graph (b) at cycle number 2, 10, 50, 100, 150, 200 and 300. (e) FESEM image of the post-cycled MnO$_2$@CNTF hybrid.*

The origin of such increase in the specific capacity over CD cycling can be inferred from comparison of the voltage profiles, as shown in **Figure 6d** for 2, 10, 50, 100, 15, 200 and 300 cycles. Up to 50 cycles the voltage profiles are similar to those obtained in rate capability measurements, characterised by prominent redox plateaus for the conversion reactions. For further cycles, the lithiation profile shows a decrease in the plateau corresponding to the reduction of manganese (region ii) at the expense of an increase in the pseudocapacitive regions (i) and (iii). Similarly, in the delithiation profiles the plateaus around 1.2 V and 2.4 V slowly disappear and the delithiation profiles during the late cycles resemble an almost straight charging profile. This indicates a clear increase in the pseudocapacitive processes compared to conversion reactions.

To understand this consecutive change in the mechanism of charge storage process, we performed 'post-mortem' analysis of the cycled electrode. Although direct observation of the active material is complicated by the presence of the SEI (**Figure S10**), there is evidence (**Figure 6d**) of the transformation of the original porous morphology of the $MnO_2$ elongated nanocrystals into a nanostructured coating around the CNT bundles, a process known as 'electrochemical milling'.[44] During the conversion reactions the elongated interconnected $MnO_2$ nanocrystals gradually change shape without overall changes in crystal structure type, but with loss of crystallinity through the formation of new vacancies and grain boundaries. As amorphization of nanostructured $LiMnO_2$ progresses, Li storage by conversion is unfavourable and replaced by intercalation and pseudo capacitive processes. The defective structure can provide additional lithium storage sites, and together with possible increase in specific surface area, produce an increase of the pseudocapacitive contribution to total capacity. Indeed, a comparison of EIS spectra of the hybrid before and after electrochemical treatment shows a significant drop of ~ 100 Ω in the charge-transfer resistance of the hybrid after cycling (see **Figure S12** and **Table ST3** in ESI).

**Properties normalised by electrode weight**

An important attribute of the hybrid material used in this work is that it already contains the current collector "built into it". Combined with the high values of specific capacity discussed above, it makes $MnO_2$/CNTF hybrids attractive electrodes. As a reference for comparison, we produced conventional electrodes of commercial $MnO_2$ loaded with conducting carbon and 20 μm thick copper current collector (comm-$MnO_2$-C@Cu, ~ 9% $MnO_2$). The rate performance of the commercial comm-$MnO_2$-C@Cu is shown in **Figure S13**, where the capacity values are normalized with respect to the total weight of the electrode. The specific capacity of the $MnO_2$/CNTF electrode is superior by at least an order of magnitude for all current densities. Taking the ratio of specific capacity of $MnO_2$/CNTF over the reference electrode, the improvement ranges from a factor of 18 at low current densities to a factor of 64 at high current density of 5 A/g (**Figure S13c**). Fully optimised electrodes would have a larger fraction of active material relative to current collector, however, this comparison helps illustrate benefits at device level of the architecture used here based on nanostructuring the active material by direct growth onto the porous, tough, CNTF current collector.

**Conclusion**

In conclusion, we have developed a flexible MnO$_2$@CNT hybrids as anode materials for high performance Li-ion storage. The nanostructured morphology of the MnO$_2$ crystals is beneficial for the hybrid to achieve high Li capacity by providing additional surface storage sites. Using detailed electrochemical analysis, we determine that at low C-rate the charge storage mechanism mostly relies on the conversion mechanism, whereas at high C rate, the contribution of pseudocapacitive process is dominant. The high surface area of this hybrid helps to achieve a specific capacity value in excess of 1100 and 500 mAh/g at a discharge current density of 25 mA/g and 5 A/g, respectively. From the in situ WAXS measurements, an irreversible phase change of ε-MnO$_2$ to LiMnO$_2$ was observed in the first cycle. Therefore, part of the irreversible capacity loss in the 1$^{st}$ galvanostatic cycle is related to the irreversible lithium consumption by the redox active phase. In subsequent galvanostatic cycles the phase changes are reversible. We found that, at a moderate discharge rate, the storage of lithium occurs possibly through an intercalation process, followed by a conversion mechanism.

The MnO$_2$@CNT hybrid has shown interesting stability profile on galvanostatic cycling, where the specific capacity gradually increases up to 200%. Our post-mortem investigations reveal that upon cycling, the interconnected porous nanocrystals of MnO$_2$ undergo 'electrochemical milling', that results in generation of new surface storage sites, drastic change in the morphology of the oxide around CNTF along with an increase in the interface area between the oxide and CNTF. All of these effects enhanced specific capacity. Finally, we found that at high current density, the free-standing MnO$_2$@CNTF hybrid can provide up to 64 times higher specific capacity compared to a reference electrodes produced with commercial MnO$_2$ on a metallic current collector.

Our results establish that use of freestanding ε-MnO$_2$@CNTF anode is highly promising for light weight, flexible, lithium ion storage devices. The architecture of nanocrystals of MnO$_2$ directly anchored on CNTF provides low electrical resistance and stability to contain volumetric expansions during intercalation and conversion processes. But ultimately, the electrochemically reversible phase for lithium storage is found to be LiMnO$_2$, rather than parent MnO$_2$. Work is in progress to investigate hybrids with other transition metal oxides and revisit the premise of reversible conversion of the initial material.

**Acknowledgements**

M.R. and J.J.V. are grateful for generous financial support provided by the European Union Seventh Framework Program under grant agreement 678565 (ERC-STEM) and 648319 (ERC-HyMAP) the Clean Sky Joint Undertaking 2, Horizon 2020 under Grant Agreement Number 738085 (SORCERER), by MINECO (RyC-2014-15115) and FotoArt-CM (S2018/NMT-4367).

N. B. acknowledges Comunidad de Madrid for the post-doctoral fellowship 2018-T2/AMB-12025. The authors thank NCD-SWEET beamline staff at ALBA Synchrotron Light Facility for assistance with synchrotron experiments.